\documentclass[aps,pra,floatfix,superscriptaddress,showpacs,twocolumn]{revtex4}
\usepackage{amsmath}
\usepackage{graphicx}

\begin{document}
\title{Some rules of good scientific writing}
\author{Dmitry Budker}
\email{budker@berkeley.edu} \affiliation{Department of Physics,
University of California, Berkeley, CA 94720-7300}

\date{\today}
\begin{abstract}
A non-native English speaking physics professor formulates obvious
yet useful rules for writing research papers.
\end{abstract}

\maketitle

\section{Introduction}
There are many volumes written about technical writing, and I
probably do not have too much original material to add to them. Yet,
whenever a student brings me a draft of their first research paper,
I invariably see an almost universal set of problems. Some of these
probably stem from the way writing is taught at school. As an
example, at school we are taught to enrich our writing by avoiding
repeating the same term and using synonyms instead. Unfortunately,
if one is writing a scientific paper, using different words for the
same object could be a disaster.

Hoping that these notes will actually be read, let us, without
further ado, present

\section{The rules}
\begin{itemize}
\item A wise man said: ``If you can abstain from
writing -- do not write!''
\item ``When in doubt -- cross it out.'' Try it; it really works
miracles!
\item The contents of a section should match its title.
\item An equation appearing in the text should never be presented
without comment, unless it is an intermediate step in a derivation.
\item All ``letters'' (i.e., variables and constants) appearing in
equations should be explicitly defined, even if seemingly obvious.
\item All references, figures, tables, and equations should be
numbered in order of appearance.
\item Sentences cannot start with an abbreviation [e.g., Fig. 1 or Eq. (2)], or with  ``So'' or
``Also.''
\item It is usually better to use past indefinite tense, for example ``it was found'' (as opposed
to present or past perfect -- ``it has or had been found''), unless
necessary.
\item Saying ``This was demonstrated by J. Doe (1905)'' is
correct, while saying ``This was demonstrated in J. Doe (1905)" is
not.
\item Things to be compared shall be presented in a similar manner
(for example, on graphs with the same scale).
\item One should avoid self-praise, for example, saying that
``interesting results were obtained." It should be up to the reader
to praise the work!
\item The reader does not know what comes next in the paper; consider what the reader should be thinking as they reach this particular point.
\item Avoid colloquial terms, for example, ``slam'' in ``The
projectile slams the target.''
\item This one is a must: read the finished manuscript!
\end{itemize}

\section{Conclusion}
These rules are quite obvious and ``common sense.'' Yet, formulating
them explicitly and keeping them in mind while writing could,
hopefully, be useful. It goes without saying, that as with most
rules, there may be exceptions.

Do we follow our own advice? Judge for yourself by checking out some
of the recent published work of our group at
\verb1http://socrates.berkeley.edu/~budker/1.

\section*{Acknowledgment}
I am grateful to all of my present and and former students and
co-authors, and, particularly, to Derek F. Jackson Kimball (now a
Professor at California State University, East Bay) for making these
rules so apparent.

\end{document}